\documentclass[ tightenlines, 
preprintnumbers, showpacs, nofootinbib,superscriptaddress,notitlepage,11pt]{revtex4-1}

\usepackage{graphicx}
\usepackage{amsmath,amsthm,amssymb}
\usepackage{mathtools}
\usepackage{color}
\usepackage{siunitx}

\usepackage{lipsum}

\newcommand{\beq}{\begin{eqnarray}}
\newcommand{\eeq}{\end{eqnarray}}
\newcommand{\nn}{\nonumber \\}

\newcommand{\Slash}[1]{{\ooalign{\hfil/\hfil\crcr$#1$}}}
\newcommand{\Tr}{\text{Tr}}
\newcommand{\phys}{\text{phys}}

\newcommand{\ave}[1]{\langle #1 \rangle}

\begin{document}

\preprint{YITP-16-133}

\title{Gluon orbital angular momentum at small-$x$ }

\author{Yoshitaka Hatta} 

\author{Yuya Nakagawa} 
\affiliation{Yukawa Institute for Theoretical Physics, Kyoto University, Kyoto 606-8502, Japan}

\author{Bowen Xiao}
\affiliation{    Key Laboratory of Quark and Lepton Physics (MOE) and Institute
of Particle Physics, Central China Normal University, Wuhan 430079, China}

\author{Feng Yuan}
\affiliation{Nuclear Science Division, Lawrence Berkeley National
Laboratory, Berkeley, CA 94720, USA}

\author{Yong Zhao}
\affiliation{Nuclear Science Division, Lawrence Berkeley National
Laboratory, Berkeley, CA 94720, USA}
\affiliation{Maryland Center for Fundamental Physics, University of Maryland, College Park, MD 20742, USA}
\affiliation{Center for Theoretical Physics, Massachusetts Institute of Technology, Cambridge, MA 02139, USA}

\begin{abstract}
We present a general analysis of the orbital angular momentum (OAM) distribution of gluons $L_g(x)$ inside the nucleon with particular emphasis on the small-$x$ region. We derive a novel operator representation of $L_g(x)$ in terms of Wilson lines and argue that it is approximately proportional to the gluon helicity distribution $L_g(x) \approx -2\Delta G(x)$ at small-$x$.  We also compute  longitudinal single spin asymmetry in exclusive diffractive dijet production in  
lepton-nucleon scattering in the next-to-eikonal approximation and show that  the asymmetry is a direct probe of the gluon helicity/OAM distribution as well as the QCD odderon exchange.  
\end{abstract}
\maketitle

\section{Introduction}

After nearly thirty years since the discovery of `spin crisis' by the EMC 
collaboration~\cite{Ashman:1987hv}, the partonic decomposition of the nucleon spin 
continues to be a fascinating research area. Among the four terms in the Jaffe-Manohar 
decomposition formula~\cite{Jaffe:1989jz}, 
\beq
\frac{1}{2}=\frac{1}{2}\Delta \Sigma + \Delta G + L_q+L_g\,, \label{1}
\eeq 
the quark helicity contribution $\Delta \Sigma$ is reasonably well constrained by the experimental data. 
The currently accepted value is $\Delta \Sigma \sim 0.30$. Over the past decade or so, there have been 
worldwide experimental efforts  to determine the gluon helicity contribution $\Delta G$ as the integral of 
the polarized gluon distribution function $\Delta G=\int_0^1 dx \Delta G(x)$. The most recent NLO global 
QCD analysis has found a nonvanishing gluon polarization in the moderate $x$ region 
$\int_{0.05}^1dx \Delta G(x)\approx 0.2^{+0.06}_{-0.07}$~\cite{deFlorian:2014yva}. 
However, uncertainties from the small-$x$ region $x<0.05$ are quite large, of order 
unity. Future experimental data from RHIC at $\sqrt{s}=510$ GeV~\cite{Adare:2015ozj} and the planned
Electron-Ion Collider (EIC)~\cite{Accardi:2012qut} are 
expected to drastically reduce these uncertainties.

In contrast to these achievements in the helicity sector, it is quite frustrating 
that very little is known about the orbital angular momentum (OAM) of quarks $L_q$ and gluons $L_g$. 
In fact, even the proper, gauge-invariant definitions of $L_{q,g}$ have long remained obscure  (see, however, 
\cite{Bashinsky:1998if}). Thanks to  recent theoretical developments, it is now understood that  
$L_{q,g}$ can be defined in a manifestly gauge invariant (albeit nonlocal) 
way~\cite{Hatta:2011zs,Hatta:2011ku}. Moreover, this construction naturally 
allows one to define, also gauge invariantly, the associated partonic distributions~\cite{Hatta:2012cs,Ji:2012ba}, 
\beq
L_{q,g}=\int_0^1 dx L_{q,g}(x)\,. 
\eeq 
A detailed analysis shows that $L_{q,g}(x)$ is sensitive to the twist-three correlations in the longitudinally polarized nucleon. 
  
Introducing the $x$-distributions $L_{q,g}(x)$ is essential for the experimental measurement of OAMs.    Just like $\Delta \Sigma$, which is the integral of the polarized quark distribution $\Delta \Sigma=\int_0^1dx \Delta q(x)$, $L_{q,g}$ can only be determined through a global analysis of the `OAM parton distributions' $L_{q,g}(x)$ extracted from various observables.  However, accessing $L_{q,g}(x)$ experimentally is quite challenging, and there has been some recent debate over whether they can be in principle related to observables in the first place \cite{Courtoy:2013oaa,Kanazawa:2014nha,Rajan:2016tlg}.  

In this paper we propose a method to experimentally measure the {\it gluon} OAM distribution $L_{g}(x)$ for  small values of $x$. This is  practically important in view of 
the abovementioned large uncertainties in $\Delta G$ from the small-$x$ region, as well as a 
strong coupling analysis \cite{Hatta:2009ra} which suggests that a significant fraction of spin 
comes from OAM at small-$x$. Together with a related proposal which focuses on the moderate-$x$ region \cite{Ji:2016jgn}, our work  represents a major step forward towards understanding the spin sum rule (\ref{1}).\footnote{
Very recently, a different observable related to the quark OAM distribution $L_q(x)$ for generic values of $x$ \cite{Bhattacharya:2017bvs} has been suggested. Moreover, the first direct computation of $L_q$ in lattice QCD simulations \cite{Engelhardt:2017miy} has appeared. }
We shall make a crucial use of the relation \cite{Lorce:2011kd,Lorce:2011ni,Hatta:2011ku} between $L_{q,g}$ and the QCD Wigner distribution  \cite{Belitsky:2003nz}, or its Fourier transform, the generalized transverse momentum dependent  distribution (GTMD) \cite{Meissner:2009ww,Hatta:2011ku,Lorce:2013pza}, which actually holds at the density level $L_{q,g}(x)$. 
Since  the gluon Wigner distribution is measurable at small-$x$ \cite{Hatta:2016dxp}, $L_{g}(x)$ should also be measurable through this relation.

 In Section II, we review the gauge invariant gluon OAM $L_g$ and its $x$-distribution $L_g(x)$. In Section III, we discuss  the said  relation between $L_g(x)$ and the gluon Wigner distribution, and prove some nontrivial identities. From Section IV on, we focus on the small-$x$ regime.   
 We derive a novel operator representation of $L_g(x)$ in terms of lightlike Wilson lines.  The operator is unusual (for those who are familiar to nonlinear small-$x$ evolution equations) as it is comprised of half-infinite  Wilson lines and covariant derivatives. We observe that exactly the same operator is relevant to the polarized gluon distribution $\Delta G(x)$ at small-$x$. This, together with the arguments in Appendix B, has led us to advocate the relation
\beq
L_g(x) \approx -2\Delta G(x)\,, \qquad (x \ll 1) \label{ap}
\eeq
which puts strong constraints on the small-$x$ behavior of $L_g(x)$ and $\Delta G(x)$ and their uncertainties.  It also suggests that the measurement of $L_g(x)$ at small-$x$ is closely related to that of $\Delta G(x)$. Based on this expectation, in Section V   
we compute  longitudinal single spin 
asymmetry 
$d\Delta \sigma=d\sigma^\rightarrow -d\sigma^\leftarrow$ 
in diffractive dijet 
production in  lepton-nucleon scattering. It turns out that the asymmetry vanishes in the leading eikonal approximation, and the first nonvanishing contributions come from the next-to-eikonal corrections. This involves precisely the OAM operator found in Section IV, and as a result, the asymmetry is directly proportional to $L_g(x)$ in certain kinematic regimes. Interestingly, the asymmetry is also proportional to the odderon amplitude in QCD. Finally, we comment on the small-$x$ evolution of $L_g(x)$ and $\Delta G(x)$ in Sec.~VI and conclude in Sec.~VII.


\section{Gluon orbital angular momentum}

In this section, we  review the gluon OAM $L_g$ and its associated parton distribution $L_g(x)$ following \cite{Hatta:2011zs,Hatta:2011ku,Hatta:2012cs}. The precise gauge invariant definition of $L_g$ is given by the nonperturbative proton matrix element
\beq
\lim_{\Delta\to 0} \langle P'S|F^{+\alpha}\overleftrightarrow{D}_{\rm pure}^iA^{\rm phys}_{\alpha}|PS\rangle = -i\epsilon^{ij}\Delta_{\perp j} S^+L_g\,, \label{def}
\eeq
where $P^\mu\approx \delta^\mu_+ P^+$ is the proton momentum and the spin vector is longitudinally polarized $S^\mu \approx \delta^\mu_+ S^+$. On the right hand side,  we keep only the linear term in the transverse momentum transfer $\Delta_\perp = P'_\perp-P_\perp$ which is assumed to be small. 
We use the notations $\overleftrightarrow{D}^\mu \equiv \frac{\partial^\mu -\overleftarrow{\partial}^\mu}{2}+igA_\mu$ and $D_{\rm pure}^\mu\equiv D^\mu-igA_{\rm phys}^\mu$. 
 $A^\mu_{\rm phys}$ is a nonlocal operator defined by \cite{Hatta:2011zs}
\beq
  A_{\pm {\rm phys}}^\mu(y) = \mp \int dz^- \theta(\pm(z^- - y^-))\widetilde{U}_{y^-,z^-}(y_\perp)F^{+\mu}(z^-,y_\perp)\,, \label{aphys}
\eeq
 where $\widetilde{U}$ is the lightlike Wilson line segment in the adjoint representation. $L_g$ does not depend on the choice of the $\pm$ sign in  
(\ref{aphys}) due to $PT$ symmetry \cite{Hatta:2011ku}. In the light-cone gauge $A^+=0$, $A_{\rm phys}^\mu=A^\mu$ and (\ref{def}) reduces to the canonical gluon OAM originally introduced  by  Jaffe and Manohar \cite{Jaffe:1989jz}.  
The operator structure (\ref{def}) was first written down in \cite{Chen:2008ag}, but the authors proposed a different $A_{\rm phys}^\mu$. We emphasize that the choice (\ref{aphys}) is unique if one identifies $\Delta G$ in (\ref{1})  with the usual gluon helicity $\Delta G$ that has been measured at RHIC and other experimental facilities.

Next we discuss the gluon OAM distributions $L_{g}(x)$ with the property\footnote{The normalization of $L_g(x)$ in (\ref{norm}) and (\ref{lg}) differs by a factor of 2 from that in   Ref.~\cite{Hatta:2012cs} where $L_g(x)$ was defined as $L_g=\int_{-1}^1dx L_g(x)=2\int_0^1dx L_g(x)$. The present choice is in parallel with the definition of $\Delta G(x)$: $\int_0^1dx \Delta G(x)=\Delta G$. }
\beq
L_g= \int_{0}^1 dx L_g(x) = \frac{1}{2}\int_{-1}^1 dx L_g(x)\,. \label{norm}
\eeq
 The $x$-distributions for the quark and gluon OAMs $L_{q,g}(x)$ have been previously introduced in  \cite{Hagler:1998kg,Harindranath:1998ve} and their DGLAP evolution equation has been derived to one-loop. However, the definition in \cite{Hagler:1998kg,Harindranath:1998ve} is not gauge invariant, and the computation of the anomalous dimensions has been performed in the light-cone gauge $A^+=0$. The gauge invariant canonical OAM distributions $L_{q,g}(x)$ have been first introduced in \cite{Hatta:2012cs}. They reduce to the previous definitions \cite{Hagler:1998kg,Harindranath:1998ve} if one takes the light-cone gauge.\footnote{There is an alternative gauge invariant  definition in  \cite{Hoodbhoy:1998yb}, but this is different from the one \cite{Hatta:2012cs} we discuss in the following. }  
 While the notion of OAM parton distributions is not yet widely known, we emphasize that they are crucial for the measurability of    OAMs. Just as  one has to measure the polarized quark and gluon distributions $\Delta q(x),\Delta G(x)$ in order to extract $\Delta \Sigma=\int_0^1 dx \Delta q(x)$ and $\Delta G=\int_0^1 dx \Delta G(x)$, any attempt to experimentally determine $L_{q,g}$ must start by  measuring its $x$-distribution $L_{q,g}(x)$.

For the gauge invariant gluon OAM (\ref{def}) with $A_{\rm phys}^\mu$ given by (\ref{aphys}), the distribution $L_g(x)$ is also gauge invariant and is defined through the relation \cite{Hatta:2012cs}
\beq
\delta(x-x')\frac{L_g(x)}{2} = \frac{M_F(x,x')}{x(x-x')} -\frac{M_D(x,x')}{x}\,, \label{lg}
\eeq 
 where $M_F$ and $M_D$ are the `F-type' and `D-type' three-gluon collinear correlators
\beq
 &&\int  \frac{dy^-  dz^-}{(2\pi)^2} e^{i xP^+y^- +i(x'-x)P^+z^-}  \langle P'S|F^{+\alpha}(0) gF^{+i}(z^-)F^{+}_{\ \alpha}(y^-)|PS\rangle  \nn 
&&= -ixP^+ \int  \frac{dy^-  dz^-}{(2\pi)^2} e^{i xP^+y^- +i(x'-x)P^+z^-}  \langle P'S|F^{+\alpha}(0) gF^{+i}(z^-)A^{\pm 
{\rm phys}}_\alpha(y^-)|PS\rangle    \nn 
&&=\epsilon^{ij}\Delta_{\perp j} S^+ M_F(x,x')+\cdots \,, \label{three}
\eeq
\beq
&& \int \frac{dy^- dz^-}{(2\pi)^2} e^{ixP^+y^- +i(x'-x)P^+z^-}  \langle P'S|F^{+\alpha}(0) \overleftrightarrow{D}^{i}(z^-)F^{+}_{\ \alpha}(y^-)|PS\rangle \nn 
&&=-ixP^+ \int \frac{dy^- dz^-}{(2\pi)^2} e^{ixP^+y^- +i(x'-x)P^+z^-}  \langle P'S|F^{+\alpha}(0) \overleftrightarrow{D}^{i}(z^-)A^{\pm {\rm phys}}_\alpha(y^-)|PS\rangle \nn 
&&   =\epsilon^{ij}\Delta_{\perp j} S^+ M_D(x,x') + \cdots\,. \label{dt}
\eeq
 (In the above, we omitted Wilson lines $\widetilde{U}$ for simplicity.)  The quark OAM distribution $L_q(x)$ can be similarly defined through the collinear quark-gluon-quark operators. 
Interestingly, although $L_{q,g}(x)$ are related to three-parton correlators which are twist-three, a partonic interpretation is possible because one of the three partons has vanishing longitudinal momentum fraction  $x-x'=0$ due to the delta function constraint in (\ref{lg}).  
After using the QCD  equations of motion, one can reveal the precise twist structure of $L_g(x)$: It can be written as the sum of the `Wandzura-Wilczek' part and the genuine twist-three part   \cite{Hatta:2012cs} 
\beq
\frac{1}{2}L_g(x) &=& \frac{x}{2}\int_x^{1} \frac{dx'}{x'^2}(H_g(x')+E_g(x'))- x\int_x^{1} \frac{dx'}{x'^2} \Delta G(x') \nonumber \\
  && +2x\int_x^{1} \frac{dx'}{x'^3} \int dX \Phi_F(X,x')  +2x\int_x^{1}dx_1 \int_{-1}^1 dx_2 \tilde{M}_F(x_1,x_2) {\mathcal P}\frac{1}{x_1^3(x_1-x_2)}
\nonumber \\
 && \qquad +2 x\int_x^{1} dx_1\int_{-1}^1 dx_2 M_F(x_1,x_2) {\mathcal P} \frac{2x_1-x_2}{x_1^3(x_1-x_2)^2}\,, \label{ma}
 \eeq
 where $H_g=xG(x)$ and $E_g$ are the gluon generalized parton distributions (GPDs) at vanishing skewness. 
$\Phi_F$ and $\tilde{M}_F$ are the quark-gluon-quark and three-gluon correlators defined similarly to (\ref{three}) (see \cite{Hatta:2012cs}  for the details). Eq.~(\ref{ma}) shows that $L_g(x)$ and $\Delta G(x)$ are related, albeit in a complicated way. Later we shall find a more direct relation between the two distributions special to the small-$x$ region.

Before leaving this section, we show the DGLAP equations for $L_{q,g}(x)$. They can be extracted from the results of the anomalous dimensions in \cite{Hagler:1998kg,Harindranath:1998ve} (see, also, \cite{Ji:1995cu}). 
\beq
\frac{d}{d\ln Q^2} \left(\begin{matrix} L_q(x) \\ L_g(x) \end{matrix}\right)= \frac{\alpha_s}{2\pi} \int_x^1 \frac{dz}{z} \left(\begin{matrix} \hat{P}_{qq}(z)  &\hat{P}_{qg}(z) & \Delta \hat{P}_{qq}(z) & \Delta \hat{P}_{qg}(z) \\ \hat{P}_{gq}(z) & \hat{P}_{gg}(z) & \Delta \hat{P}_{gq}(z) & \Delta \hat{P}_{gg}(z) \end{matrix}\right) \left(\begin{matrix} L_q(x/z) \\ L_g(x/z) \\ \Delta q(x/z) \\ \Delta G(x/z) \end{matrix}\right)\,, \label{d1}
\eeq
\beq
&&\hat{P}_{qq}(z)=C_F \left(\frac{z(1+z^2)}{(1-z)_+} + \frac{3}{2}\delta(1-z) \right)\,, \\
&&\hat{P}_{qg}(z) =n_f  z(z^2+(1-z)^2)\,, \\
&&\hat{P}_{gq}(z)= C_F(1+(1-z)^2)\,, \\
&&\hat{P}_{gg}(z)= 6\frac{(z^2-z+1)^2}{(1-z)_+} +  \frac{\beta_0}{2}\delta(z-1)\,, \\
&&\Delta \hat{P}_{qq}(z)=C_F (z^2-1)\,, \\
&&\Delta \hat{P}_{qg}(z)= n_f (1-3z+4z^2-2z^3)\,, \\
&&\Delta \hat{P}_{gq}(z)= C_F (-z^2+3z-2)\,, \\
&&\Delta \hat{P}_{gg}(z)=6 (z-1)(z^2-z+2)\,,
\eeq
where $C_F=\frac{N_c^2-1}{2N_c}=\frac{4}{3}$, $n_f$ is the number of flavors and $\beta_0=11-\frac{2n_f}{3}$. 
For completeness and a later use, we also note the DGLAP equation for the helicity distributions
\beq
\frac{d}{d\ln Q^2} \left(\begin{matrix} \Delta q(x) \\ \Delta G(x) \end{matrix}\right)= \frac{\alpha_s}{2\pi} \int_x^1 \frac{dz}{z} \left(\begin{matrix} \Delta P_{qq}(z) & \Delta P_{qg}(z) \\  \Delta P_{gq}(z) & \Delta P_{gg}(z) \end{matrix}\right) \left(\begin{matrix}   \Delta q(x/z) \\ \Delta G(x/z) \end{matrix}\right)\,, \label{d2}
\eeq
\beq
&&\Delta P_{qq}(z)=C_F \left(\frac{1+z^2}{(1-z)_+} + \frac{3}{2}\delta(1-z) \right)\,, \\
&&\Delta P_{qg}(z)= \frac{n_f}{2}(2z-1)\,, \\
&&\Delta P_{gq}(z)= C_F (2-z)\,, \\
&&\Delta P_{gg}(z)=6 \left(\frac{1}{(1-z)_+}-2z+1\right) +  \frac{\beta_0}{2}\delta(z-1)\,.
\eeq

\section{OAM and the Wigner distribution}

The original definition (\ref{lg}) is technical and does not immediately invoke its physical meaning as the OAM. Fortunately, there exists an equivalent and very intuitive definition of $L_g(x)$ in terms of the Wigner distribution. The gluon Wigner distribution is defined  as 
\beq
xW(x,q_\perp,b_\perp,S)&=&2\int\frac{dz^-d^2z_\perp}{{(2\pi)}^3P^+}\int\frac{d^2 \Delta_\perp}{{(2\pi)}^2}e^{-ixP^+z^-+iq_\perp\cdot z_\perp} \nonumber \\
&&\times\left\langle P+\tfrac{\Delta_\perp}{2},S\left|{\rm Tr}F^{+i}\left(b_\perp+\tfrac{z}{2}\right)F^{+i}\left(b_\perp-\tfrac{z}{2}\right)\right|P-\tfrac{\Delta_\perp}{2},S\right\rangle\,,  \label{wi}
\eeq
where the trace is in the fundamental representation. 
It is convenient to also consider the Fourier transform of the Wigner distribution with respect to $b_\perp$, namely,  the generalized transverse momentum dependent distribution (GTMD)  \cite{Meissner:2009ww,Hatta:2011ku,Lorce:2013pza}
\beq
xW(x,q_\perp,\Delta_\perp,S)&=&\int d^2b_\perp W_g(x,q_\perp,b_\perp,S)e^{i\Delta_\perp\cdot b_\perp} \nn
 &=&4\int\frac{d^3x d^3y}{(2\pi)^3}e^{-ixP^+(x^--y^-)+iq_\perp\cdot(x_\perp-y_\perp)+i\frac{\Delta_\perp}{2}\cdot(x_\perp+y_\perp)} \langle {\rm Tr} F^{+i}(x)F^{+i}(y)\rangle\,, \label{gt}
\eeq 
 where $\langle\cdots\rangle\equiv\frac{\langle P+\frac{\Delta_\perp}{2},S|\cdots|P-\frac{\Delta_\perp}{2},S \rangle}{\langle P,S|P,S \rangle}$. 
In (\ref{wi}) and (\ref{gt}), we have to specify the configuration of Wilson lines to make the nonlocal operator $F(x)F(y)$ gauge invariant. There are two interesting choices for this \cite{Bomhof:2007xt,Dominguez:2011wm}. 
One is the Weizs\"acker-Williams (WW) type
\begin{align}
\ave{\Tr F^{+i}(x)F^{+i}(y)}\to\ave{\Tr F^{+i}(x)U_\pm(x,y)F^{+i}(y)U_\pm(y,x)} \,, \label{eq:WW}
\end{align}
and the other is the dipole type
\begin{align}
\ave{\Tr  F^{+i}(x)F^{+i}(y)}\to\ave{\Tr F^{+i}(x)U_-(x,y)F^{+i}(y)U_+(y,x)} \,, \label{eq:dip}
\end{align}
where $U_\pm(x,y)\equiv U_{x^-,\pm \infty}(x_\perp)U_{x_\perp,y_\perp}(\pm \infty)U_{\pm \infty,y^-}(y_\perp)$ is a staple-shaped Wilson line in the fundamental representation. 
We denote the corresponding  distributions as $W_\pm$ and $W_\text{dip}$, respectively.

The Wigner distribution describes the phase phase distribution of gluons with transverse momentum $q_\perp$ and impact parameter $b_\perp$. Their cross product $b_\perp \times q_\perp$ classically represents the orbital angular momentum.  It is thus natural to define $L_g$ as  \cite{Hatta:2011ku}
\beq
L_{g}&\equiv& \int_{-1}^1 dx \int d^2b_\perp d^2q_\perp\ \epsilon_{ij} b^i_\perp q^j_\perp W_\pm (x,q_\perp,b_\perp) \nn 
&=& -i\int_{-1}^1 dx \int d^2q_\perp\ \epsilon^{ij}q_\perp^j\lim_{\Delta_\perp\to 0}\frac{\partial}{\partial\Delta^i_\perp}W_\pm(x,q_\perp,\Delta_\perp) \,, \label{oamdef}
\eeq
where our default choice is the WW-type Wigner distribution because it is consistent with a partonic interpretation. 
One can check that (\ref{oamdef}) agrees with (\ref{def}), with the $\pm$ sign taken over to that in (\ref{aphys}). 
$W$ has the following spin-dependent structure    
\begin{align}
W(x,q_\perp,\Delta_\perp,S)=i\frac{S^+}{P^+}\epsilon^{ij} \Delta^i_\perp q_\perp^j  \bigl( f(x,|q_\perp|)+i\Delta_\perp \cdot q_\perp h(x,|q_\perp|) \bigr) +\cdots\,.    \label{w0} 
\end{align}
Substituting this into (\ref{oamdef}), one finds 
\begin{align}
L_g=\lambda \int_{-1}^1  dx \int  d^2q_\perp\ q^2_\perp f(x,|q_\perp|)\,, \label{mom}
\end{align}
 where 
$\lambda=\frac{S^+}{P^+}=\pm 1$ is the helicity of the proton.

The result (\ref{mom}), together with a similar relation for the quark OAM, is by now well established \cite{Lorce:2011kd,Lorce:2011ni,Hatta:2011ku}.  We now discuss this relation at the level of the $x$-distribution.  
Since (\ref{oamdef}) involves an integration over $x$, it is tempting to identify the integrand with $L_g(x)$ 
\beq
L_g(x) &=& 2\int d^2b_\perp d^2q_\perp\ \epsilon_{ij} b^i_\perp q^j_\perp W_\pm(x,q_\perp,b_\perp) \nn 
 &=& -2i\int d^2q_\perp\ \epsilon^{ij}q_\perp^j\lim_{\Delta_\perp\to 0}\frac{\partial}{\partial\Delta^i_\perp}W_\pm (x,q_\perp,\Delta_\perp)  \,. \label{lx}
\eeq
(The factor of 2 is because $\int_{-1}^1dx = 2\int_0^1dx$.) 
It turns out that this exactly agrees with $L_g(x)$ defined in  (\ref{lg}).  The proof was essentially given in \cite{Hatta:2012cs}  for the quark OAM distribution $L_q(x)$. The generalization to the gluon case is straightforward, and this is outlined in Appendix \ref{aa}. Here we prove another nontrivial fact  that $L_g(x)$'s  defined through the WW and dipole Wigner distribution are identical for all values of $x$. Namely, 
\beq
\int \!d^2b_\perp d^2q_\perp \epsilon^{ij}b_\perp^i q_\perp^j W_\pm(x,b_\perp,q_\perp) = \int \!d^2b_\perp d^2q_\perp \epsilon^{ij}b_\perp^i q_\perp^j W_{\rm dip}(x,b_\perp,q_\perp) \,.
\label{wign} 
\eeq 
The proof goes as follows.  Consider the part that involves $q_\perp$; $\int d^2q_\perp q_\perp^j W$. For the WW-type Wigner, this is evaluated as 
\beq
&&\int d^2q_\perp q_\perp^j\int\frac{d^2z_\perp}{(2\pi)^2}e^{iq_\perp\cdot z_\perp}\Tr F^{+i}\left(\tfrac{z}{2}\right)U_\pm F^{+i}\left(-\tfrac{z}{2}\right)U_\pm^\dagger=i\lim_{z_\perp\to 0}\frac{\partial}{\partial z_\perp^j}\left(\Tr F^{+i}\left(\tfrac{z}{2}\right)U_\pm F^{+i}\left(-\tfrac{z}{2}\right)U_\pm^\dagger\right) \notag \\
&&=\frac{1}{2}\Tr \left[ F^{+i}\left(\tfrac{z^-}{2}\right)(i\overleftarrow{D}_jU-iUD_j)F^{+i}\left(-\tfrac{z^-}{2}\right)U^\dagger\right]  \nonumber \\
&&\qquad +\frac{1}{2}\Tr\left[\left[F^{+i},gA_{\pm\phys}^j \right]\left(\tfrac{z^-}{2}\right)UF^{+i}\left(-\tfrac{z^-}{2}\right)U^\dagger\right] -\frac{1}{2}\Tr \left[ F^{+i}\left(\tfrac{z^-}{2}\right)U\left[F^{+i},gA_{\pm\phys}^j\right]\left(-\tfrac{z^-}{2}\right)U^\dagger\right] \nn 
&& = \frac{1}{2}\Tr\left[ F^{+i}\left(\tfrac{z^-}{2}\right)(i\overleftarrow{D}^{\rm pure}_jU-iUD^{\rm pure}_j)F^{+i}\left(-\tfrac{z^-}{2}\right)U^\dagger \right]
\,, \label{eq:WW moment}
\eeq
where we only show the relevant operator structure and suppress the arguments of Wilson lines $U$ which should be obvious from gauge invariance. 
The same type of calculation for the dipole Wigner distribution gives
\beq
&&\int d^2q_\perp q_\perp^j\int\frac{d^2z_\perp}{(2\pi)^2}e^{iq_\perp\cdot z_\perp}\Tr F^{+i}\left(\tfrac{z}{2}\right)U_- F^{+i}\left(-\tfrac{z}{2}\right)U_+^\dagger \nonumber \\
&&=\frac{1}{2}\Tr \left[ F^{+i}\left(\tfrac{z^-}{2}\right)(i\overleftarrow{D}_jU-iUD_j)F^{+i}\left(-\tfrac{z^-}{2}\right)U^\dagger\right] \nn 
&& \qquad  +\frac{1}{2}\Tr\left[g(F^{+i}A_{-\phys}^j-A_{+\phys}^jF^{+i})\left(\tfrac{z^-}{2}\right)UF^{+i}\left(-\tfrac{z^-}{2}\right)U^\dagger\right]
\nonumber \\
&& \qquad -\frac{1}{2}\Tr\left[ F^{+i}\left(\tfrac{z^-}{2}\right)Ug(F^{+i}A_{+\phys}^j-A_{-\phys}^jF^{+i})\left(-\tfrac{z^-}{2}\right)U^\dagger\right]. \label{eq:dip moment}
\eeq
Taking the plus sign in (\ref{eq:WW moment}) (the minus sign leads to the same conclusion) and subtracting (\ref{eq:dip moment}), we obtain 
\begin{align}
&i\lim_{z_\perp\to 0}\frac{\partial}{\partial z_\perp^j}\left(\Tr F^{+i}\left(\tfrac{z}{2}\right)U_+ F^{+i}\left(-\tfrac{z^-}{2}\right)U_+^\dagger\right)
-i\lim_{z_\perp\to 0}\frac{\partial}{\partial z_\perp^j}\left(\Tr F^{+i}\left(\tfrac{z}{2}\right)U_- F^{+i}\left(-\tfrac{z}{2}\right)U_+^\dagger\right) \notag \\
&=\frac{1}{2}\Tr \left[F^{+i}\left(A_{+\text{phys}}-A_{-\text{phys}}\right)\left(\tfrac{z^-}{2}\right)UF^{+i}\left(-\tfrac{z^-}{2}\right)U^\dagger \right]\notag \\
&\qquad +\frac{1}{2}\Tr \left[F^{+i}\left(\tfrac{z^-}{2}\right)U\left(A_{+\text{phys}}-A_{-\text{phys}}\right)F^{+i}\left(-\tfrac{z^-}{2}\right)U^\dagger\right] \notag \\
&=-\int dy^-\Tr \left[F^{+i}\left(\tfrac{z^-}{2}\right)U_{\frac{z^-}{2},y^-}F^{+i}(y^-)U_{y^-,-\frac{z^-}{2}}F^{+i}\left(-\tfrac{z^-}{2}\right)U_{-\frac{z^-}{2},\frac{z^-}{2}}\right] \label{eq:diff WW +}\,.
\end{align}
The question is whether the nonforward matrix element $\langle ...\rangle$ of the operator (\ref{eq:diff WW +}) contains the structure $i\frac{S^+}{P^+}\epsilon^{ij}\Delta_\perp^i \delta L(x)$. If so, the function $\delta L$ would contribute to the difference $L^{WW}_g(x)-L^{\rm dip}_g(x)$.  However, this is impossible as one can easily see by applying the $PT$ transformation to the matrix element. Under $PT$, $F^{\mu\nu}\to -F^{\mu\nu}$, and one obtains an identity 
\begin{align}
&i\frac{S^+}{P^+}\epsilon^{ij}\Delta_\perp^i \delta L(x)=-i\frac{-S^+}{P^+}\epsilon^{ij}(-\Delta_\perp^i)\delta L(x)\,,
\end{align}
which immediately gives $\delta L(x)=0$.

The above proof is crucial for the measurability of $L_g(x)$.  While $L_g(x)$ is naturally defined by the WW-type Wigner distribution, the dipole Wigner distribution has a better chance to be measured in experiments \cite{Hatta:2016dxp}. Below we only consider $W_{\rm dip}$ and omit the subscript.

\section{Small-$x$ regime}

Our discussion so far has been general and valid for any value of $x$. From now on, we focus on the small-$x$ regime. In this section we derive a novel operator representation of $L_g(x)$ and point out its unexpected relation  to the polarized gluon distribution $\Delta G(x)$. 

\subsection{Leading order}

In order to study the properties of the (dipole) Wigner distribution at small-$x$, as a first step we approximate   
 $e^{  -ixP^+(x^- - y^-)}\approx 1$ in (\ref{gt}). We shall refer to this as the eikonal approximation.  We then use the identity 
\beq
\partial_i U(x_\perp) 
&=& -ig\int_{-\infty}^\infty dx^- U_{\infty, x^-} F^{+i}(x) U_{x^-,-\infty}  -igA^i(\infty,x_\perp)U(x_\perp) +igU(x_\perp)A^i(-\infty,x_\perp)\,,
\label{las}
\eeq 
where $U(x_\perp)\equiv U_{\infty,-\infty}(x_\perp)$ and do integration by parts. This leads us to    
 \cite{Hatta:2016dxp}
\beq
W (x,\Delta_\perp,q_\perp,S)    \approx W_0 (x,\Delta_\perp,q_\perp)   
   =  \frac{4N_c}{ x g^2 (2\pi)^3} \left(q_\perp^2 - \tfrac{\Delta_\perp^2}{4}\right)  F(x,\Delta_\perp,  q_\perp)\,, \label{der}
\eeq
where $F$ is the Fourier transform of the so-called dipole S-matrix
\beq
 F(x,\Delta_\perp, q_\perp) \equiv  \int d^2x_\perp d^2y_\perp e^{iq_\perp \cdot (x_\perp-y_\perp)+ i(x_\perp + y_\perp)\cdot \frac{\Delta_\perp}{2}} 
 \left\langle \frac{1}{N_c}  {\rm Tr} \left[U(x_\perp) U^\dagger(y_\perp)\right]  \right\rangle\,. \label{dipF} 
\eeq
The last two terms in (\ref{las}) have been canceled against the terms which come from the derivative of the  transverse gauge links connecting $x_\perp$ and $y_\perp$ at $x^-=\pm \infty$ (not shown in (\ref{dipF}) for simplicity). The $x$-dependence of $F$ arises from the quantum evolution of the dipole operator ${\rm Tr}U(x_\perp)U^\dagger(y_\perp)$.  
To linear order in $\Delta_\perp$, we can parameterize $F$ as 
\beq
 F(x,\Delta_\perp,q_\perp) 
 = P(x,\Delta_\perp,q_\perp) + iq_\perp \cdot \Delta_\perp O(x,|q_\perp| )\,. \label{p}
\eeq
The imaginary part $O$ comes from the so-called odderon operator \cite{Hatta:2005as,Zhou:2016rnt}. It is important to notice that $F$ cannot depend on the longitudinal spin $S^+$, and therefore, $W_0$ cannot have the structure (\ref{w0}). This follows from $PT$ symmetry which dictates that
\beq
\left\langle P+\tfrac{\Delta}{2},S\left| {\rm Tr}[U(x_\perp)U^\dagger(y_\perp)]\right|P-\tfrac{\Delta}{2},S\right \rangle = 
\left\langle P-\tfrac{\Delta}{2},-S\left| {\rm Tr}[U(-x_\perp)U^\dagger(-y_\perp)]\right|P+\tfrac{\Delta}{2},-S\right \rangle \,, \nonumber
\eeq
 so that  $W_0(x,q_\perp,\Delta_\perp,S)=W_0(x,-q_\perp,-\Delta_\perp,-S)$. 
Therefore,  it is impossible to access any information about spin and OAM in the eikonal approximation. This is actually expected on physical grounds. At high energy, spin effects are suppressed by a factor of $x$ (or inverse energy) compared to the `Pomeron' contribution as represented by the first term $P$ in (\ref{p}).\footnote{The situation is different when the spin is transversely polarized. In this case, $F$ can have the structure  $\epsilon_{ij}S_{\perp}^i q_\perp^j$, and the corresponding amplitude has been dubbed the `spin-dependent odderon' \cite{Zhou:2013gsa}. While this is subleading compared to the leading Pomeron term $P$, it is suppressed only by a fractional power $x^\alpha$ with $\alpha \sim 0.3$. }  

\subsection{First subleading correction}

In order to be sensitive to the spin and OAM effects, we have to go beyond the eikonal approximation. By taking into account the second term in the expansion $e^{-ixP^+(x^- - y^-)} = 1-ixP^+(x^- - y^-)+\cdots$ and writing $W=W_0+\delta W$ accordingly, we find 
\beq
 \delta W (x,\Delta_\perp,q_\perp,S) &=&-\frac{4P^+}{ g (2\pi)^3 }\int d^2x_\perp d^2y_\perp e^{iq_\perp \cdot (x_\perp-y_\perp)+ i(x_\perp + y_\perp)\cdot \frac{\Delta_\perp}{2} }  \nn 
&& \times \Biggl\{ 
  \int_{-T}^T dx^- ( x^- +T) \frac{\partial}{\partial y_\perp^i} \left\langle {\rm Tr} \left[U_{T,x}F^{+i}(x)U_{x,-T} U^\dagger(y_\perp)  \right] \right\rangle    \nn 
&& + \int_{-T}^T dy^- (y^- +T) \frac{\partial}{\partial x_\perp^i} \left\langle  {\rm Tr} \left[U(x_\perp) U_{-T,y}F^{+i}(y)U_{y,T} \right] \right\rangle \Biggr\} \nn 
&=&\frac{4P^+}{g^2(2\pi)^3}\int d^2x_\perp d^2y_\perp e^{i(q_\perp +\frac{\Delta_\perp}{2})\cdot x_\perp + i(-q_\perp+ \frac{\Delta_\perp}{2})\cdot y_\perp }   \nn 
&& \times \Biggl\{ 
  \int_{-T}^T dz^-  \left(q_\perp^i-\tfrac{\Delta^i_\perp}{2}\right)\left\langle  {\rm Tr} \left[  U_{Tz^-} (x_\perp)\overleftarrow{D}_i  U_{z^--T} (x_\perp) U^\dagger(y_\perp)  \right] \right\rangle  \nn 
&& + \int_{-T}^T dz^- \left(q_\perp^i +\tfrac{\Delta^i_\perp}{2} \right) \left\langle  {\rm Tr} \left[U(x_\perp)  U_{-Tz^-}(y_\perp) D_i U_{z^-T} (y_\perp)  \right] \right\rangle \Biggr\} \,. \label{yo} 
\eeq
The first equality is obtained by splitting $x^- - y^- = x^- + T -(y^- +T)$ where $T$ is eventually sent to infinity. In the second equality 
we write $x^- + T = \int_{-T}^{x^-} dz^-$ and switch the order of integrations between $\int dx^-$ and $\int dz^-$. 

In contrast to  $W_0$,  $\delta W$ can have the structure (\ref{w0}): From PT symmetry, one can show that
 $\delta W(x,q_\perp,\Delta_\perp,S)=-\delta W(x,-q_\perp,-\Delta_\perp,-S)$.\footnote{More generally, in the Taylor expansion of the phase factor $e^{-ixP^+(x^--y^-)}$, the odd terms in $x$ can contribute to the OAM.}
 The most  general parameterization of  the near-forward matrix element in (\ref{yo}) is, to linear order in $\Delta_\perp$ and $S^+$, 
\begin{align}
&\frac{4P^+}{g^2(2\pi)^3}\int d^2x_\perp d^2y_\perp\ e^{i(q_\perp+\frac{\Delta_\perp}{2})\cdot x_\perp+i(-q_\perp+\frac{\Delta_\perp}{2})\cdot y_\perp} \int dz^-\ \ave{\Tr[U_{\infty,z^-}(x_\perp)\overleftarrow{D}_iU_{z^-,-\infty}(x_\perp)U^\dagger(y_\perp)]} \nonumber \\
&=-i\frac{S^+}{2P^+}\epsilon_{ij}\Biggl\{\Biggl(q_\perp^j+\tfrac{\Delta_\perp^j}{2}\Biggr)f(x,|q_\perp|)+\Biggl(q_\perp^j-\tfrac{\Delta_\perp^j}{2}\Biggr)g(x,|q_\perp|)+q_\perp^j\Delta_\perp\cdot q_\perp A(x,|q_\perp|)\Biggr\} \nonumber \\
&\quad -\frac{S^+}{2P^+}\epsilon_{ij}\Biggl\{\Biggl(q_\perp^j+\tfrac{\Delta_\perp^j}{2}\Biggr)B(x,|q_\perp|)+\Biggl(q_\perp^j-\tfrac{\Delta_\perp^j}{2}\Biggr)C(x,|q_\perp|)-2q_\perp^j\Delta_\perp\cdot q_\perp h(x,|q_\perp|)\Biggr\} +\cdots \,. \label{eq:1 para'}
\end{align}
\begin{align}
&\frac{4P^+}{g^2(2\pi)^3}\int d^2x_\perp d^2y_\perp\ e^{i(q_\perp+\frac{\Delta_\perp}{2})\cdot x_\perp+i(-q_\perp+\frac{\Delta_\perp}{2})\cdot y_\perp} \int dz^-\ave{\Tr[U(x_\perp)U_{-\infty,z^-}(y_\perp)D_iU_{z^-,\infty}(y_\perp)]} \nonumber \\
&=i\frac{S^+}{2P^+}\epsilon_{ij}\Biggl\{\Biggl(q_\perp^j-\tfrac{\Delta_\perp^j}{2}\Biggr)f(x,|q_\perp|)+\Biggl(q_\perp^j+\tfrac{\Delta_\perp^j}{2}\Biggr)g(x,|q_\perp|)-q_\perp^j\Delta_\perp\cdot q_\perp A(x,|q_\perp|) \Biggr\}\nonumber \\
&\quad -\frac{S^+}{2P^+}\epsilon_{ij}\Biggl\{\Biggl(q_\perp^j-\tfrac{\Delta_\perp^j}{2}\Biggr)B(x,|q_\perp|)+\Biggl(q_\perp^j+\tfrac{\Delta_\perp^j}{2}\Biggr)C(x,|q_\perp|)+2q_\perp^j\Delta_\perp\cdot q_\perp h(x,|q_\perp|) \Biggr\}+\cdots\,. \label{eq:1 para cc}
\end{align}
(\ref{eq:1 para cc}) is obtained from (\ref{eq:1 para'}) by applying the $PT$ transformation. 
We recognize the functions $f$ and $h$ that appear in (\ref{w0}), the former is related to the OAM as in (\ref{mom}). The other real-valued functions $g, A, B, C$ do not contribute to the Wigner distribution.  
Integrating both sides over $q_\perp$, we obtain the following  sum rules
\beq
\int d^2q_\perp \left( f- g+q_\perp^2 A\right)=0\,, \qquad
 \int d^2q_\perp \left( B-C-2q_\perp^2 h\right)=0 \,. \label{fo}
\eeq

 Eq.~(\ref{eq:1 para'}) uncovers a novel representation of the OAM distribution at small-$x$ in terms of an unusual Wilson line operator in which the covariant derivative $D_i$ is inserted at an intermediate time $z^-$. Such operators do not usually appear in the context of high energy evolution.  In the next section we shall see  that this structure is related to the next-to-eikonal approximation. Here we point out that the same operator is relevant to the polarized gluon distribution $\Delta G(x)$. This elucidates an unexpected relation between $\Delta G(x)$ and $L_g(x)$.   

Let us define the `unintegrated' (transverse momentum dependent) polarized gluon distribution $\Delta G(x,q_\perp)$ as
\begin{align}
&ix\Delta G(x,q_\perp)\frac{S^+}{P^+}\equiv 2\int \frac{d^2z_\perp dz^-}{(2\pi)^3P^+}e^{-ixP^+z^-+iq_\perp\cdot z_\perp}\left\langle PS\left|\epsilon_{ij}\Tr F^{+i}\left(\tfrac{z}{2}\right)U_- F^{+j}\left(-\tfrac{z}{2}\right)U_+\right|PS\right\rangle \notag \\
& \qquad =4\int\frac{d^3x d^3y}{(2\pi)^3} e^{-ixP^+(x^--y^-) +iq_\perp \cdot (x_\perp-y_\perp)}\frac{\langle PS| \epsilon_{ij} {\rm Tr} \left[F^{+i}(x)U_- F^{+j}(y)U_+\right]|PS\rangle }{\langle PS|PS\rangle}\,, \label{uni}
\end{align}
such that $\int d^2q_\perp \Delta G(x,q_\perp) = \Delta G(x)$ and $\int_0^1dx \Delta G(x)=\Delta G$. 
Note that (\ref{uni}) is a forward matrix element $\Delta_\perp=0$. 
Using the same approximation as above, we obtain the following representation at small-$x$
\begin{align}
  i\Delta G(x,q_\perp)\frac{S^+}{P^+}&= \frac{4P^+}{g^2(2\pi)^3}\int d^2x_\perp d^2y_\perp e^{i q_\perp \cdot( x_\perp - y_\perp) }   \notag \\
& \times \epsilon_{ij} \Biggl\{  q_\perp^j
  \int_{-\infty}^\infty dz^-  \left\langle  {\rm Tr} \left[  U_{\infty z^-} (x_\perp)\overleftarrow{D}_i  U_{z^--\infty} (x_\perp) U^\dagger(y_\perp)  \right] \right\rangle  \notag \\ 
& \qquad \qquad + q_\perp ^i \int_{-\infty}^\infty dz^-  \left\langle  {\rm Tr} \left[U(x_\perp)  U_{-\infty z^-}(y_\perp) D_j U_{z^-\infty} (y_\perp)  \right] \right\rangle \Biggr\} \,,
\end{align} 
or equivalently,
\beq
&& \Delta G(x,q_\perp)\frac{S^+}{P^+ } \\
&&= \frac{8P^+}{g^2(2\pi)^3} \epsilon_{ij}  q_\perp^j{\mathfrak Im} \left[\int d^2x_\perp d^2y_\perp e^{i q_\perp \cdot( x_\perp - y_\perp) } 
  \int_{-\infty}^\infty dz^-  \left\langle  {\rm Tr} \left[  U_{\infty z^-} (x_\perp)\overleftarrow{D}_i  U_{z^--\infty} (x_\perp) U^\dagger(y_\perp)  \right] \right\rangle \right] \,. \nonumber
\eeq
Substituting (\ref{eq:1 para'}), we find
\begin{align}
\Delta G(x)&= -\int d^2q_\perp  q_\perp^2 (f(x,|q_\perp|) + g(x,|q_\perp|)) \nn
&=-\frac{1}{2}L_g(x) -\int d^2q_\perp q^2_\perp g(x,|q_\perp|)\,.   \label{eq:G=L}
\end{align}
This is a rather surprising result. From (\ref{ma}), one can argue that if $\Delta G(x)$ shows a power-law behavior at small-$x$, $\Delta G(x) \sim x^{-\alpha}$, the OAM distribution grows with the same exponent $L_g(x)\sim x^{-\alpha}$. Eq.~(\ref{eq:G=L}) imposes a strong constraint on the respective prefactors, and the relation is preserved by the small-$x$ evolution  because both $L_g(x)$ and $\Delta G(x)$ are governed by the same operator. Moreover, in Appendix \ref{ab} we present three different arguments which indicate that  $|f|\gg |g|$. If this is true,  a very intriguing  relation emerges 
\beq
L_g(x)\approx -2\Delta G(x) \,. \label{sur}
\eeq
As mentioned in the introduction, reducing the huge uncertainty in $\Delta G$ from the small-$x$ region $x<0.05$ \cite{deFlorian:2014yva} is a pressing issue in QCD spin physics. Eq.~(\ref{sur}) suggests that, if the integral $\int_0^{0.05} dx \Delta G(x)$ turns out to be sizable in future, one should expect an even larger contribution from the gluon OAM in the same $x$-region which reverses the sign of the net gluon angular momentum
\beq
\int^{0.05}_0 dx \Delta G(x) + \int^{0.05}_0 dx L_g(x) \approx  -\int_0^{0.05} dx \Delta G(x)\,. \label{e50}
\eeq
This has profound implications on the spin sum rule (\ref{1}). In particular, it challenges the idea that $\Delta \Sigma$ and $\Delta G$ alone can saturate the sum rule. There must be OAM contributions. 

Eq.~(\ref{sur}) is reminiscent of a similar relation observed in the large-$Q^2$ asymptotic scaling behavior of  the components in the spin decomposition formula  Eq.~(\ref{1})
\cite{Ji:1995cu}. To one-loop order, 
\begin{eqnarray}
\Delta\Sigma(t)&=&{\rm const}. \,,\\
L_q(t)&=&-\frac{1}{2}\Delta\Sigma +\frac{1}{2}\frac{3n_f}{16+3n_f} \ ,\\
\Delta G(t)&=& -\frac{4\Delta\Sigma}{\beta_0}+\frac{t}{t_0}\left(\Delta G_0+\frac{4\Delta\Sigma}{\beta_0}\right)\ ,\label{e53}\\
L_g(t)&=&-\Delta G(t)+\frac{1}{2}\frac{16}{16+3n_f} \ ,\label{e54}
\end{eqnarray}
where $t=\ln\left(Q^2/\Lambda_{QCD}^2\right)$
and we have neglected the subleading terms at large-$Q^2$.
 $\Delta G_0$ represents the gluon helicity contribution
at some initial scale $t_0$. From these equations, we find that the large negative
gluon orbital angular momentum would cancel out the gluon helicity contribution
if the latter is large and positive. It is interesting to see how this behavior 
imposes a constraint on the small-$x$ contribution to $\Delta G$ and $L_g$ when we apply 
Eq.~(\ref{e50}) as the initial condition. 
The scale evolution of $L_g(x)$ and $\Delta G(x)$  can be an important agenda for 
the future electron-ion collider~\cite{Accardi:2012qut} where one of the primary goals is to 
investigate the sum rule (\ref{1}).

\section{Single spin asymmetry in diffractive dijet production} 

In this section, we calculate longitudinal single spin asymmetry in  
 forward dijet production in exclusive diffractive lepton-nucleon scattering. 
As observed recently \cite{Hatta:2016dxp}, in this process  one can probe the  gluon 
Wigner distribution at small-$x$ (see also \cite{Altinoluk:2015dpi}) and its characteristic 
angular correlations.  Here we show that the same process, with the proton being 
longitudinally polarized, is directly sensitive to the function $f(x,q_\perp)$.

\subsection{Next-to-eikonal approximation}

Exclusive diffractive forward dijet production in $ep$ collisions  has been extensively studied in the literature mostly in the BFKL framework  \cite{Nikolaev:1994cd,Bartels:1996ne,Bartels:1996tc,Braun:2005rg,Boussarie:2016ogo}, and more recently in the color glass condensate framework \cite{Altinoluk:2015dpi,Hatta:2016dxp}.  
We work in the so-called dipole frame where the left-moving virtual photon with virtuality $Q^2$ 
splits into a $q\bar{q}$ pair and scatters off the right-moving proton. The proton emerges elastically with momentum transfer $\Delta_\perp$. The $q\bar{q}$ pair is detected in the forward region (i.e., at large negative rapidity) as two jets with the total transverse 
momentum $k_{1\perp}+k_{2\perp}=-\Delta_\perp$ and the relative momentum 
$\frac{1}{2}(k_{2\perp}-k_{1\perp})=P_\perp$.

In the eikonal approximation and for the transversely polarized virtual photon, the 
amplitude is proportional to \cite{Altinoluk:2015dpi,Hatta:2016dxp}
\beq 
&\propto&  \int d^2x_\perp  d^2 y_\perp e^{-ik_{1\perp}\cdot x_\perp - ik_{2\perp} \cdot y_\perp} \left\langle\frac{1}{N_c}{\rm Tr}[U(x_\perp)U^\dagger(y_\perp)] \right\rangle  \frac{\varepsilon K_1(\varepsilon r_\perp)}{2\pi} \frac{r^i_\perp }{r_\perp}  \nn  
&=& i\int  \frac{d^2q_\perp}{(2\pi)^2} \frac{P^i_\perp -q^i_\perp}{(P_\perp-q_\perp)^2+\varepsilon^2} F(\Delta_\perp,q_\perp),
 \label{gene}
\eeq
where $r_\perp = x_\perp-y_\perp$ and $\varepsilon^2=z(1-z)Q^2$.  $z$ (or $1-z$) is the longitudinal momentum fraction of the virtual photon energy $q^-$ carried by  the quark (or antiquark). 

As we already pointed out, (\ref{gene}) cannot depend on spin. Our key observation is that the next-to-eikonal corrections to (\ref{gene}) include exactly the same matrix element as   (\ref{eq:1 para'}) and is therefore sensitive to the gluon OAM function $f$. Going beyond the eikonal approximation, we generalize (\ref{gene}) as 
\beq
 \int d^2x_\perp d^2x'_\perp d^2 y_\perp d^2y'_\perp e^{-ik_{1\perp}\cdot x_\perp - ik_{2\perp} \cdot y_\perp}  \left\langle\frac{1}{N_c}{\rm Tr}[U(x_\perp,x'_\perp)U^\dagger(y_\perp,y'_\perp)] \right\rangle  \frac{\varepsilon K_1(\varepsilon r'_\perp)}{2\pi} \frac{r'^i_\perp }{r'_\perp}\,, \label{go} 
\eeq
where we allow the quark and antiquark to change their transverse coordinates during propagation.  $U(x_\perp,x'_\perp)$ is essentially the Green function and can be determined as follows. 

 Consider the propagation of a quark with energy $k^-=zq^-$ in the background field $A^+$, $A_\perp^i$. The Green function satisfies the equation\footnote{For a quark, there is an extra term in the equation at ${\cal O}(1/k^-)$ which depends on the gamma matrices $\Slash D \Slash D = D^2 + \frac{g}{2}\sigma_{\mu\nu}F^{\mu\nu}$. We neglect this term because it gives vanishing contribution to the physical cross section to ${\cal O}(1/k^-)$ since  ${\rm Tr} \, \sigma_{\mu\nu}=0$.    }  
\beq
\left[ i\frac{\partial}{\partial x^-} + \frac{1}{2k^-}D^2_{x_\perp} - gA^+(x^-,x_\perp) \right] G_{k^-}(x^-,x_\perp,x'^-,x'_\perp) =i\delta(x^- - x'^-)\delta^{(2)}(x_\perp-x'_\perp)\,.
\eeq   
To zeroth order in $1/k^-$, the solution is 
\beq
G^0 _{k^-}(x^-,x_\perp,x'^-,x'_\perp)  = \theta(x^- - x'^-)\delta^{(2)}(x_\perp-x'_\perp) \exp\left(-ig\int^{x^-}_{x'^-}dz^- A^+(z^-,x_\perp)\right)\,.
\eeq 
This is the eikonal approximation. 
Writing $G=G^0 + \delta G$, we find the equation for $\delta G$
\beq
\left[ i\frac{\partial}{\partial x^-}  - gA^+(x^-,x_\perp) \right]\delta G +   \frac{1}{2k^-}D^2_{x_\perp}  G^0=0\,.
\eeq
This can be easily solved as
\beq
\delta G(x^-,x_\perp,x'^-,x'_\perp) =\frac{i}{2k^-} \theta(x^-- x'^-)\int_{x'^-}^{x^-} dz^- U_{x^-z^-}(x_\perp) D^2_{x_\perp}  \delta^{(2)}(x_\perp-x'_\perp) U_{z^-x'^-}(x'_\perp)\,.
\eeq
We thus obtain the desired propagator 
\beq
U(x_\perp,x'_\perp) &\equiv & G_{k^-}(\infty,x_\perp, -\infty, x'_\perp) \nn 
&=& U(x_\perp) \delta^{(2)}(x_\perp-x'_\perp) +\frac{i}{2k^-}\int^\infty_{-\infty}
dz^- U_{\infty z^-}(x_\perp)D^2_{x_\perp}
 \delta^{(2)}(x_\perp-x'_\perp) U_{z^--\infty}(x'_\perp)\,.
\label{propdef}
\eeq
In (\ref{go}), we need the Fourier transform of $U(x_\perp,x'_\perp)$ 
\beq
&&\int d^2x_\perp e^{-ik_\perp \cdot x_\perp } U(x_\perp,x'_\perp)\nn 
&& =e^{-ik_\perp \cdot x'_\perp} \left(U(x'_\perp) + \frac{i}{2k^-} \int_{-\infty}^\infty dz^- 
U_{\infty z^-}(x'_\perp) (\overleftarrow{D}^2_{x'_\perp} -k_\perp^2 -2ik^i_\perp \overleftarrow{D}_{x'^i_\perp}) U_{z^- -\infty}(x'_\perp) \right)\,. \label{al}
\eeq
If we ignore $A_\perp$, (\ref{al}) agrees with the result of \cite{Altinoluk:2014oxa,Altinoluk:2015gia} to the order of interest, although equivalence is not immediately obvious.\footnote{ Note that the $k_\perp^2$ term comes from the expansion of the on-shell phase factor 
\beq
e^{-ik^+\int dz^-}= \exp\left(-i\frac{k_\perp^2}{2k^-}\int dz^-\right) \approx 1-i\frac{k_\perp^2}{2k^-}\int dz^-\,. \nonumber
\eeq
This term is proportional to the leading term and can be dropped since it does not give any spin-dependence. 
}
 Clearly, $A_\perp$ is important for the result to be gauge invariant (covariant). The last term in (\ref{al}), when substituted into (\ref{go}), gives the same operator as in (\ref{eq:1 para'}). In addition, (\ref{al}) contains the operator  $U_{\infty,z^-}\overleftarrow{D}^2_{x_\perp}U_{z^-,-\infty}$ which we did not encounter in the previous section.  
 However, the matrix element of this operator does not require new functions. To see this, we write down the general parameterization  to linear order in $\Delta_\perp$
\begin{align}
&\frac{4P^+}{g^2(2\pi)^3}\int d^2x_\perp d^2y_\perp e^{iq_\perp\cdot(x_\perp-y_\perp)+i\frac{\Delta_\perp}{2}\cdot(x_\perp+y_\perp)}\int dz^- \ave{\Tr[U_{\infty,z^-}(x_\perp)\overleftarrow{D}^2_{x_\perp}U_{z^-,-\infty}(x_\perp)U^\dagger(y_\perp)]} \notag \\
&=\bigl( \kappa(x,|q_\perp|) + i\eta(x,|q_\perp|)   \bigr) \frac{S^+}{P^+}\epsilon^{ij}q_\perp^i\Delta_\perp^j+\cdots\,,\label{cc}
\end{align}
\begin{align}
&\frac{4P^+}{g^2(2\pi)^3}\int d^2x_\perp d^2y_\perp e^{iq_\perp\cdot(x_\perp-y_\perp)+i\frac{\Delta_\perp}{2}\cdot(x_\perp+y_\perp)}\int dz^- \ave{\Tr[U_{\infty,z^-}(x_\perp)\overrightarrow{D}^2_{x_\perp}U_{z^-,-\infty}(x_\perp)U^\dagger(y_\perp)]} \notag \\
&=-\bigl( \kappa(x,|q_\perp|) + i\eta(x,|q_\perp|)   \bigr) \frac{S^+}{P^+}\epsilon^{ij}q_\perp^i\Delta_\perp^j+\cdots\,,\label{cc2}
\end{align}
where $\kappa$, $\eta$ are real.  (\ref{cc}) and (\ref{cc2}) are related by $PT$ symmetry. 
By integrating by parts in (\ref{cc}) twice, we can replace the operator $U_{\infty,z^-}\overleftarrow{D}^2_{x_\perp}U_{z^-,-\infty}$ with   a linear combination of $U_{\infty,z^-}\overrightarrow{D}^2_{x_\perp}U_{z^-,-\infty}$  and the surface terms. The latter can depend on spin through the operator
\beq
i \left(q^i_\perp + \frac{\Delta^i_\perp}{2}\right) 
U_{\infty,z^-}\overleftarrow{D}_{x_\perp^i}U_{z^-,-\infty}\,,
\eeq
as in  (\ref{eq:1 para'}). We thus obtain an identity
\beq
 \kappa + i\eta   = -( \kappa + i\eta ) + \frac{g}{2} -i\frac{C}{2}\,,
\eeq
 and therefore, 
\begin{align}
\kappa(x,|q_\perp|)=\frac{1}{4}g(x,|q_\perp|)\,, \qquad 
\eta(x,|q_\perp|)=-\frac{1}{4}C(x,|q_\perp|)\,.
\end{align}

\subsection{Calculation of the asymmetry}

We are now ready to compute the longitudinal single spin asymmetry. 
\beq
\frac{d\Delta \sigma }{dy_1d^2k_{1\perp} dy_2 d^2k_{2\perp} } \equiv 
\frac{d\sigma^{\lambda=+1} }{dy_1d^2k_{1\perp} dy_2 d^2k_{2\perp} }  -\frac{d\sigma^{\lambda=-1} }{dy_1d^2k_{1\perp} dy_2 d^2k_{2\perp} }\,,  
\eeq
where $y_1$, $y_2$ are the rapidities of the two jets. 
Our strategy is the following. We first substitute (\ref{al}) into (\ref{go}) and use the parameterizations (\ref{eq:1 para'}) and (\ref{cc}) for the resulting matrix elements. We then square the amplitude and keep only the linear terms in $S^+/k^-$.   
The leading eikonal contribution has both the real and imaginary parts from the Pomeron and odderon exchanges, respectively
\beq
\int d^2q_\perp \frac{P_\perp ^i -q_\perp^i}{(P_\perp -q_\perp)^2+\varepsilon^2} \left(P(\Delta_\perp,q_\perp) + i\Delta_\perp \cdot q_\perp O(q_\perp) \right)\,. \label{dom}
\eeq
The next-to-eikonal contribution of order $1/k^-$ also contains both  real and imaginary parts  as shown in  (\ref{eq:1 para'}) and (\ref{cc}). When squaring the amplitude, we see that the terms linear in $S^+$ arises from the interference between the leading and next-to-eikonal contributions. It turns out that the odderon $O$ interferes with the imaginary terms in (\ref{eq:1 para'}) which in particular include the OAM function $f$, while the Pomeron $P$ interferes with the real terms  in (\ref{eq:1 para'})  which we are not interested in. The problem is that, on general grounds, one expects that the Pomeron amplitude $P$ is numerically larger than the odderon amplitude $O$, and this can significantly reduce the sensitivity to the OAM function. We avoid this problem by focusing on the following two kinematic regions 
\beq
P_\perp \gg q_\perp, Q\,, \qquad Q\gg q_\perp, P_\perp\,.
\eeq 
($q_\perp$ here means the typical values of $q_\perp$ within the support of the functions $P$ and $O$.) In this limit, the Pomeron contributionin (\ref{dom})  drops out because 
\beq
\int d^2q_\perp P(\Delta_\perp,q_\perp) = 0, \qquad \int d^2q_\perp q_\perp^i P(\Delta_\perp,q_\perp)=0\,,
\eeq
for $\Delta_\perp \neq 0$. The first integral vanishes because the $q_\perp$-integral sets the dipole size $r_\perp=x_\perp-y_\perp$ to be zero so that $U(x_\perp)U^\dagger(x_\perp)=1$. Thus the integral becomes proportional to the delta function $\delta^{(2)}(\Delta_\perp)$. The second relation follows from the symmetry $P(\Delta_\perp,q_\perp)=P(\Delta_\perp,-q_\perp)$. 
On the other hand, the odderon contribution survives in this limit because, for example,  
\beq
\int d^2q_\perp q_\perp^i \Delta_\perp \cdot q_\perp O(q_\perp) = \frac{\Delta_\perp^i}{2}\int d^2q_\perp q_\perp^2 O(q_\perp)\,.
\label{od}
\eeq
We can thus approximate, when $P_\perp\gg q_\perp,Q$,
\begin{align}
\int d^2q_\perp \frac{P_\perp ^i -q_\perp^i}{(P_\perp -q_\perp)^2+\varepsilon^2} \left(P(\Delta_\perp,q_\perp) + i\Delta_\perp \cdot q_\perp O(q_\perp) \right) \approx i \left(-\frac{\Delta_\perp^i}{2P_\perp^2} + \frac{P_\perp^i P_\perp \cdot \Delta_\perp}{P_\perp^4}\right)\int d^2q_\perp q_\perp^2 O(q_\perp)\,. \label{4}
\end{align}
 A similar result follows in the other limit $Q\gg q_\perp, P_\perp$.   (\ref{4}) is to be multiplied by the next-to-eikonal amplitude which reads
\beq
&&
\int \frac{d^2q_\perp}{(2\pi)^2} \frac{ P_\perp^i-q_\perp^i}{(P_\perp-q_\perp)^2 + \epsilon_f^2}\int d^2x'_\perp  d^2 y'_\perp e^{i(q_\perp + \frac{\Delta_\perp}{2}) \cdot x'_\perp + i(-q_\perp + \frac{\Delta_\perp}{2}) \cdot y'_\perp} \nn 
&& \qquad \times \int dz^- \Biggl\langle   \frac{1}{k_1^-} {\rm Tr} U_{\infty z^-}(x'_\perp) \left( k_{1\perp}^j  \overleftarrow{D}'_j +\frac{i}{2}\overleftarrow{D}'^2\right) U_{z^-,-\infty}(x'_\perp) U^\dagger(y'_\perp) 
\nn
&& \qquad \qquad \qquad   - \frac{1}{k^-_2} {\rm Tr} U(x'_\perp) U_{-\infty z^-}(y'_\perp) \left(k_{2\perp}^j D_j' + \frac{i}{2}\overrightarrow{D}'^2\right) U_{z^-\infty}(y'_\perp)  \Biggr\rangle \nn
 &&= \frac{i\lambda}{4}\frac{g^2(2\pi)^3}{4P^+} \int \frac{d^2q_\perp}{(2\pi)^2} \frac{ P_\perp^i-q_\perp^i}{(P_\perp-q_\perp)^2 + \varepsilon^2} 
 \nn
&& \quad  \times \Biggl[ 
\left(\frac{1}{k_1^-}+\frac{1}{k_2^-}\right) \epsilon_{jk} \left((f-g)  P_\perp^j \Delta_\perp^k -(f+g)q_\perp^j  \Delta_\perp^k + 2A\Delta_\perp \cdot q_\perp P_\perp^j q_\perp^k +2\kappa q_\perp^j \Delta_\perp^k \right) \nn
&&  \qquad \qquad +\left( \frac{1}{k_1^-}-\frac{1}{k_2^-}\right) \epsilon_{jk} \left( 2(f+g) P_\perp^j q_\perp^k + A\Delta_\perp \cdot q_\perp \Delta_\perp^jq_\perp^k  \right) 
 \Biggr]+\cdots\,,
  \label{we}
\eeq
where we kept only the imaginary part. 
Here, $k_1^-= zq^-$, $k_2^- =(1-z) q^-$ and $k_{1\perp}= -\frac{\Delta_\perp}{2}-P_\perp$, and $k_{2\perp}=-\frac{\Delta_\perp}{2}+P_\perp$. We then expand the integrand in powers of $1/P_\perp$ or $1/Q$ and perform the angular integral over $\phi_q$. Consider, for definiteness,  the large-$P_\perp$ limit. At first sight, the dominant contribution  comes from the ${\cal O}(1)$ terms proportional to $\frac{P_\perp^i P_\perp^j}{P_\perp^2} (f-g)$ and $\frac{P_\perp^i P_\perp^j}{P_\perp^2} A$. However, after the $\phi_q$-integral they cancel  exactly due to  the sum rule (\ref{fo}). Thus the leading terms are ${\cal O}(1/P_\perp)$ and actually come from the last line of (\ref{we}) which can be evaluated as 
\beq
\approx \frac{i\lambda}{4}\frac{g^2(2\pi)^3}{4P^+} \left( \frac{1}{k_1^-}-\frac{1}{k_2^-}\right)\frac{\epsilon_{ij}P_\perp^j}{P_\perp^2} \int \frac{ d^2q_\perp}{(2\pi)^2} q_\perp^2(f+g) =-\frac{i\lambda \alpha_s(1-2z)}{32P^+q^-}\Delta G(x)\frac{\epsilon_{ij}P_\perp^j}{P_\perp^2}\nn \approx \frac{i\lambda \alpha_s(1-2z)}{64P^+q^-}L_g(x)\frac{\epsilon_{ij}P_\perp^j}{P_\perp^2}
\,,  \label{fin}
\eeq
where we used (\ref{eq:G=L}) and (\ref{sur}). 
Multiplying (\ref{fin}) by (\ref{4}) and restoring the prefactor, we finally arrive at
\beq
\frac{d\Delta \sigma }{dy_1d^2k_{1\perp} dy_2 d^2k_{2\perp} } 
&\approx & 
 4\pi^4 \alpha_s N_c\alpha_{em} x\sum_q e_q^2 \delta(x_{\gamma^*}-1)(1-2z)(z^2+(1-z)^2) \nn 
&&  \qquad \times \frac{\Delta_\perp}{ P_\perp^3Q^2} \sin \phi_{P\Delta} \left\{ \begin{matrix} -2 \Delta G(x) \\ L_g(x) \end{matrix} \right\}  \int d^2q_\perp q^2_\perp O (x, q_\perp)
\,, \label{rema}
\eeq

where $\phi_{P\Delta}$ is the azimuthal angle between $P_\perp$ and $\Delta_\perp$ and $e_q$ is the electric charge of the massless quark  in units of $e$. We also used $x=\frac{Q^2}{2P^+q^-}$. $z$ is fixed by the dijet kinematics as 
\beq
z=\frac{|k_{1\perp}|e^{y_1}}{|k_{1\perp}| e^{y_1}+|k_{2\perp}| e^{y_2}}\, .
\eeq 
 In the other limit  $Q\gg q_\perp,P_\perp$, the cross section reads
 \beq
\frac{d\Delta \sigma }{dy_1d^2k_{1\perp} dy_2 d^2k_{2\perp} } 
&\approx & 
4 \pi^4 \alpha_s N_c\alpha_{em} x\sum_f e_f^2 \delta(x_{\gamma^*}-1)(1-2z) \frac{z^2+(1-z)^2}{z^2(1-z)^2}
 \nn 
&&  \qquad \times \frac{P_\perp \Delta_\perp}{ Q^6} \sin \phi_{P\Delta} \left\{ \begin{matrix} -2 \Delta G(x) \\ L_g(x) \end{matrix} \right\} \int d^2q_\perp q^2_\perp O (x, q_\perp)
\,. \label{q}
\eeq
The terms neglected in (\ref{rema}) and (\ref{q}) are suppressed by powers of $1/P_\perp$ and $1/Q$, respectively. 

The above results have been obtained for the transversely polarized virtual photon.  In fact, the whole contribution from the longitudinally polarized virtual photon is subleading. The only difference in the longitudinal  photon case is the integral kernel 
\beq
\int d^2q_\perp \frac{P_\perp ^i -q_\perp^i}{(P_\perp -q_\perp)^2+\varepsilon^2} \to \int d^2q_\perp \frac{Q}{(P_\perp -q_\perp)^2+\varepsilon^2}\,.
\eeq
 Proceeding as before, we find that the contribution from the longitudinal photon to $\Delta \sigma$ is suppressed by factors $1/P_\perp^3$ and $1/Q^2$ compared to (\ref{rema}) and (\ref{q}), respectively.   

We thus find that the asymmetry is directly proportional to $\Delta G(x)$. On the basis of  (\ref{sur}), we may also say that it is proportional to $L_g(x)$.  Previous direct measurements of $\Delta G(x)$ (or rather, the ratio  $\langle \Delta G(x)/G(x)\rangle$ averaged over a limited interval of $x$) in DIS are based on longitudinal {\it double} spin asymmetry \cite{Adolph:2012ca,Adolph:2012vj}.
 In general, longitudinal  single spin asymmetry vanishes in QCD due to parity. Here, however, we get a nonzero result because we measure the correlation between two particles (jets) in the final state.  The experimental signal of this is the $\sin \phi_{P\Delta}$ angular dependence. 
This is distinct from the leading angular dependence of the dijet cross section  $\cos 2\phi_{P\Delta}$ \cite{Hatta:2016dxp} which has been canceled in the difference  $d\Delta \sigma=d\sigma^{\lambda=1} -d\sigma^{\lambda=-1}$. 

Notice that the asymmetry vanishes at the symmetric point $z=1/2$ and the product $(1-2z)\sin \phi_{P\Delta}$ is invariant under the exchange of two jets $z\leftrightarrow 1-z$ and $k_{1\perp}\leftrightarrow k_{2\perp}$.  
 Subleading corrections to (\ref{rema}) include terms proportional  to $\sin 2\phi_{P\Delta}$ without a prefactor $1-2z$.
These are consequences of parity.
Compared to $\sin \phi_{P\Delta}$,  $\sin 2\phi_{P\Delta}$ has an extra zero at $\phi_{P\Delta}=\pi/2$, or equivalently, $|k_{1\perp}| = |k_{2\perp}|$. When $z=1/2$ and $|k_{1\perp}| = |k_{2\perp}|$, the two jets cannot be distinguished. Therefore, the $\lambda=\pm 1$ cross sections are exactly equal by parity and the asymmetry vanishes.  This argument can be generalized to higher Fourier components. 
The most general form of  longitudinal single spin asymmetry consistent with parity  is 
\begin{align}
\frac{d\Delta\sigma}{dy_1d^2{k_1}_\perp dy_2d^2{k_2}_\perp}&=\sum_{n=0}^\infty c_n(z,Q,|P_\perp|,|\Delta_\perp|)\sin(2n+1)\phi_{P_\perp\Delta_\perp} \notag \\
&+\sum_{n=1}^\infty d_n(z,Q,|P_\perp|,|\Delta_\perp|)\sin 2n\phi_{P_\perp\Delta_\perp}\,,
\end{align}
where $c_n(z=\frac{1}{2},Q,|P_\perp|,|\Delta_\perp|)=0$. 

It is very interesting that the  measurement of (\ref{rema}) also establishes the odderon exchange in QCD which has long  evaded detection despite many attempts in the past \cite{Ewerz:2003xi}. The connection between odderon and (transverse) single spin aymmetries has been previously discussed in the literature \cite{Hagler:2002nf,Kovchegov:2012ga,Zhou:2013gsa,Boer:2015pni}. However, the observable and the mechanism considered in this work are new.  To estimate the cross section quantitatively, the integral  $\int d^2q_\perp q_\perp^2 O(x,q_\perp)$ should be evaluated  using models including the QCD evolution effects. Importantly, theory predicts \cite{Bartels:1999yt,Chachamis:2016ejm} that  $O(x,q_\perp)$ has no or very weak dependence on $x$ in the linear BFKL regime. This will make the extraction of the $x$-dependence of $\Delta G(x)$ easier.   

\section{Comments on the small-$x$ evolution equation}

 The appearance of half-infinite Wilson line operators is quite unusual in view of  the standard approaches to high energy QCD evolution which only deal with infinite Wilson lines $U_{\infty,-\infty}$.  At the moment, little is known about the small-$x$ evolution of these  operators. Still, we can formally write down the evolution equation by assuming that the soft gluon emissions only affect Wilson lines at the end points $x^-=\pm \infty$ \cite{Ferreiro:2001qy}. 
 Defining ${\cal O}_{x_\perp}\equiv \int dz^- U_{\infty z^-}(x_\perp)\overleftarrow{D}U_{z^-,-\infty}(x_\perp)$ and 
using the technique illustrated in \cite{Ferreiro:2001qy},
 we obtain 
\begin{eqnarray}
&&\frac{\partial }{\partial \ln 1/x} \textrm{Tr}\left[ {\cal O}_{x_\perp} U^\dagger_{y_\perp}\right] \nn
&&=  \frac{\alpha_s N_c}{2\pi^2} \int d^2 z_\perp \frac{(x_\perp -y_\perp)^2}{(x_\perp -z_\perp)^2(z_\perp -y_\perp)^2}\left\{  \frac{1}{N_c
}\textrm{Tr}\left[{\cal O}_{x_\perp}U^\dagger_{z_\perp} \right] \textrm{Tr}\left[ U_{z_\perp}U^\dagger_{y_\perp}\right]- \textrm{Tr}\left[  {\cal O}_{x_\perp}U^\dagger_{y_\perp} \right]\right\} \nonumber \\
&&\quad +\frac{\alpha_s N_c }{2\pi^2} \int d^2 z_\perp \frac{(x_\perp -z_\perp)\cdot (y_\perp -z_\perp)}{(x_\perp -z_\perp)^2(z_\perp -y_\perp)^2}\left\{ \frac{1}{N_c} 
\textrm{Tr}\left[  {\cal O}_{x_\perp}U^\dagger_{x_\perp}\right]\textrm{Tr}\left[ U_{x_\perp}U^\dagger_{y_\perp}\right]-\textrm{Tr}\left[  {\cal O}_{x_\perp}U^\dagger_{y_\perp}\right] \right\} \nonumber \\
&&\quad +\frac{\alpha_s  N_c}{2\pi^2} \int d^2 z_\perp \left[\frac{(x_\perp -z_\perp)\cdot (y_\perp -z_\perp)}{(x_\perp -z_\perp)^2(z_\perp -y_\perp)^2}-\frac{1}{(x_\perp-z_\perp)^2}\right] \nonumber \\
 && \qquad \times \left\{ \frac{1}{N_c} 
\textrm{Tr}\left[  {\cal O}_{x_\perp}U^\dagger_{z_\perp}\right]\textrm{Tr}\left[ U_{z_\perp}U^\dagger_{y_\perp}\right] -\frac{1}{N_c} \textrm{Tr}\left[ U_{x_\perp}U^\dagger_{z_\perp}\right]
\textrm{Tr}\left[ U_{z_\perp}U_{x_\perp}^\dagger {\cal O}_{x_\perp} U^\dagger_{y_\perp} \right] \right\} . \label{right}
 \end{eqnarray}

One can show that
\beq
{\cal O}_{x_\perp} U_{x_\perp}^\dagger = \int dz^- U_{\infty,z^-}\overleftarrow{D}U_{\infty,z^-}^\dagger\,,
\eeq
is an element of the Lie algebra of SU(3). Therefore, its trace, which appears on the second line of the right hand side of (\ref{right}), vanishes. 
Note that  there is no singularity at $z_\perp=y_\perp$ and $z_\perp=x_\perp$. The latter can be seen from the identity 
\beq
 \frac{(x_\perp -y_\perp)^2}{(x_\perp -z_\perp)^2(z_\perp -y_\perp)^2}+2\frac{(x_\perp -z_\perp)\cdot (y_\perp -z_\perp)}{(x_\perp -z_\perp)^2(z_\perp -y_\perp)^2}-\frac{1}{(x_\perp-z_\perp)^2} = \frac{1}{(y_\perp-z_\perp)^2}\,.
\eeq
 The above equation is similar  to the ones discussed in \cite{Kovchegov:2015pbl,fabio}. In particular,  ${\cal O}_{x_\perp}$ and the next-to-eikonal operators in (\ref{propdef}) are possibly related to the operator $V^{pol}$ introduced, but unspecified in \cite{Kovchegov:2015pbl}. If this is the case, the small-$x$ behavior of  $L_g(x)$ and $\Delta G(x)$ is related to that of the $g_1(x)$ structure function or the polarized quark distribution $\Delta q(x)$. This issue certainly deserves further study.

\section{Conclusions}

In this paper, we first presented a general analysis of the OAM gluon distribution $L_g(x)$ by making several clarifications regarding its definition and properties. We then focused on the small-$x$ regime and derived a novel operator representation for $L_g(x)$ 
in terms of half-infinite Wilson lines $U_{\pm\infty, z^-}$ and the covariant derivatives $D^i$.
It turns out that the exactly the same operators describe the polarized gluon distribution $\Delta G(x)$. Based on this, we have argued that $L_g(x)$ and $\Delta G(x)$ are proportional to each other with the relative coefficient $-2$. Moreover, the small-$x$ evolution of these distributions can be related to that of the polarized quark distribution. These observations shed new light on  the nucleon spin puzzle.      


We have also pointed out that the same operator shows up in the next-to-eikonal approximation \cite{Altinoluk:2014oxa,Altinoluk:2015gia}. This allows us to relate the helicity and OAM distributions to observabes. We have shown that single longitudinal spin asymmetry in diffractive dijet production in
 lepton-nucleon collisions is a sensitive probe of the gluon OAM  in certain kinematic regimes.  

The large-$x$ region, on the other hand, requires a different treatment and the first result has been recently reported in \cite{Ji:2016jgn} to which our work is complementary. Probing the quark OAM $L_q$ seems more difficult, but there are interesting recent developments \cite{Engelhardt:2017miy,Bhattacharya:2017bvs}. Together they open up  ways to access the last missing pieces in the spin decomposition formula (\ref{1}), and we propose to explore this direction at the EIC.

\section*{ Acknowledgements}
We thank Guillaume Beuf, Edmond Iancu, Cedric Lorc\'{e}  for discussions. 
This material is based upon work supported by the Laboratory Directed Research and Development 
Program of Lawrence Berkeley National Laboratory, the U.S. Department of Energy, 
Office of Science, Office of Nuclear Physics, under contract number 
DE-AC02-05CH11231 and DE-FG02-93ER-40762. Y.~Z. is also supported by the U.S. Department of Energy, Office of Science, Office of Nuclear Physics, within the framework of the TMD Topical Collaboration.

\appendix

\section{Equivalence of $L_g(x)$ defined in (\ref{lg}) and (\ref{lx})}
\label{aa}

In this appendix we show that $L_g(x)$'s defined in (\ref{lg}) and (\ref{lx}) are equivalent. 
We  rewrite the operator in (\ref{dt}) as 
\begin{eqnarray}
&& F^{+\alpha}(0)\widetilde{U}_{0z}\overleftrightarrow{D}^i(z^-)\widetilde{U}_{zy}A_\alpha^{\pm {\rm phys}}(y^-) \nn
&& = \frac{1}{2}F^{+\alpha}(0) \biggl(
\widetilde{U}_{0y}D^i(y^-)  + i\int^{z^-}_{y^-}d\omega^- \widetilde{U}_{0\omega}gF^{+i}(\omega^-)\widetilde{U}_{\omega y}   \label{app} \\ 
 && \qquad \qquad \qquad \qquad  -\overleftarrow{D}^i(0)\widetilde{U}_{0y}+ i\int^{z^-}_0 d\omega^- \widetilde{U}_{0\omega}gF^{+i}(\omega^-)\widetilde{U}_{\omega y} \biggr) A_\alpha^{\pm {\rm phys}}(y^-) \nn
&&= F^{+\alpha}(0) \biggl(
\frac{\widetilde{U}_{0y}D^i_{\rm pure}(y^-) 
  -\overleftarrow{D}^i_{\rm pure}(0)\widetilde{U}_{0y}}{2}  \mp i\int d\omega^- \theta(\pm (\omega^--z^-)) \widetilde{U}_{0\omega}gF^{+i}(\omega)\widetilde{U}_{\omega y} \biggr) A_\alpha^{\pm {\rm phys}}(y^-) \nonumber \,.
\end{eqnarray}
To obtain the second equality we need to split the integral
\beq
\int^{z^-}_{y^-} d\omega^-= \mp \int_{-\infty}^\infty d\omega^- \theta(\pm(\omega^- -z^-)) \pm \int^{\infty}_{- \infty}d\omega^- \theta(\pm (\omega^-- y^-))\,,
\eeq
and similarly for $\int^{z^-}_0d\omega^-$. Substituting (\ref{app}) into (\ref{dt}) and comparing with (\ref{lg}), we find
\beq
\epsilon^{ij}\Delta_{\perp j}S^+ L_g(x) = i\int \frac{dy^-}{2\pi}e^{ixP^+y^-}\langle PS|F^{+\alpha}(0)(\widetilde{U}_{0y}D^i_{\rm pure}-\overleftarrow{D}^i_{\rm pure}\widetilde{U}_{0y})A_\alpha^{\pm {\rm phys}}(y^-)|PS\rangle\,. \label{a3}
\eeq
Integrating over $x$, we recover (\ref{def}). 
(\ref{a3}) exactly agrees with the OAM defined through the WW-type Wigner distribution (\ref{lx}) as one can see from   (\ref{eq:WW moment}). 

\section{Arguments for $L_g(x)\approx -2\Delta G(x)$ }
\label{ab}

In this appendix, we discuss the function $g(x,q_\perp)$ defined in (\ref{eq:1 para'}) which accounts for the difference  between $L_g(x)$ and $\Delta G(x)$ according to (\ref{eq:G=L}). While we cannot make rigorous statements about this nonperturbative function, we give  three arguments that $g(x,q_\perp)$ is suppressed relative to the OAM function $f(x,q_\perp)$. 

\subsection{$g$ in the parton model}

First, let us evaluate $f$ and $g$ in the `parton model'. Namely, we compute the matrix element 
\beq
&&\int d^2x_\perp d^2y_\perp e^{i(q_\perp +\frac{\Delta_\perp}{2})\cdot x_\perp + i(-q_\perp+ \frac{\Delta_\perp}{2})\cdot y_\perp } \nn   
 && \qquad \times \int_{-\infty}^\infty dz^-  \left\langle P+\tfrac{\Delta}{2}\left| {\rm Tr} \left[  U_{\infty z^-}(x_\perp) \overleftarrow{D}_i U_{z^--\infty} (x_\perp) U^\dagger(y_\perp)  \right] \right|P-\tfrac{\Delta}{2} \right\rangle\,,
\eeq
in one-loop perturbation theory by replacing the external proton state with a superposition of single quark states as 
\beq
\langle P+\Delta/2|....|P-\Delta/2\rangle_{\rm proton} \to \sum_f \int \frac{d\xi}{\xi} \phi_f(\xi,\Delta_\perp) \langle \xi P+\Delta/2| ....|\xi P-\Delta/2\rangle_f\,,
\label{quark}
\eeq 
 where $\phi_f(\xi,\Delta_\perp)$ is a weight function and $f$ is the quark flavor. 
Expanding the operator to quadratic order in $A^\mu$, we find that the $S^+$-dependence  can arise only from the terms  
\beq
\sim \int dz^-\int dw^- \langle A_i(z^-,x_\perp)A^+(w^-,y_\perp)\rangle_f\,, \label{a1}
\eeq
and 
\beq
\sim \int dz^-\int dw^- \langle A_i(z^-,x_\perp)A^+(w^-,x_\perp)\rangle_f\,. \label{a2}
\eeq

For quark matrix elements, (\ref{a1}) can be evaluated as, up to a normalization factor,
\beq
 &&\frac{1}{(q_\perp+\frac{\Delta_\perp}{2})^2(q_\perp-\frac{\Delta_\perp}{2})^2} \bar{u}' \Biggl[\xi P^+(\gamma^+\gamma^-\gamma_i +\gamma_i\gamma^-\gamma^+) +q_{\perp j}(\gamma^+\gamma^j\gamma_i-\gamma_i\gamma^j\gamma^+)\Biggr] u \nn 
&& \quad \sim \frac{1}{\left(q_\perp+\frac{\Delta_\perp}{2}\right)^2\left(q_\perp-\frac{\Delta_\perp}{2}\right)^2}\epsilon_{ij}\left(q^j_\perp+\tfrac{\Delta^j_\perp}{2}\right)\xi S^+\,,  \label{jj}
\eeq
where we used $\bar{u}'\gamma^iu \approx  i\epsilon^{ij}\Delta_j \frac{S^+}{P^+}$ and computed only the imaginary part. 
As for (\ref{a2}), we get 
\beq
-\delta^{(2)}\left(q_\perp-\tfrac{\Delta_\perp}{2}\right) \int d^2k_\perp \frac{1}{\left(k_\perp+\frac{\Delta_\perp}{2}\right)^2\left(k_\perp-\frac{\Delta_\perp}{2}\right)^2} \epsilon_{ij} \frac{\Delta_\perp^j}{2}\xi S^+ \,. \label{su}
\eeq
Because of the delta function, the factor $\Delta^j_\perp/2$ in (\ref{su}) can be  replaced by $\frac{1}{2}(q_\perp^j + \tfrac{\Delta_\perp^j}{2})$. This  shows that $g=A=0$, and
\beq
f\propto  \frac{1}{\left(q_\perp+\frac{\Delta_\perp}{2}\right)^2\left(q_\perp-\frac{\Delta_\perp}{2}\right)^2}   - \delta^{(2)}\left(q_\perp-\frac{\Delta_\perp}{2}\right) \int d^2k_\perp \frac{1}{\left(k_\perp+\frac{\Delta_\perp}{2}\right)^2\left(k_\perp-\frac{\Delta_\perp}{2}\right)^2}\,.
\eeq
 It is easy to check that  the sum rule (\ref{fo}) is satisfied  to this order. 
This result suggests that $g$ is a higher order effect, suppressed at least by a factor of $\alpha_s$ compared to $f$.

 \subsection{Nonperturbative argument}

Next we give a more formal argument from another perspective.
Let us simplify the notation as  
\begin{align}
&{\cal O}_i(x_\perp)=\int dz^-\ U_{\infty,z^-}(x_\perp)\overleftarrow{D}_iU_{z^-,-\infty}(x_\perp)\,, \\
&{\cal O}'(y_\perp)=U^\dagger(y_\perp)\,,
\end{align}
and consider the matrix element
\begin{align}
&\int d^2x_\perp d^2y_\perp e^{i\left(q_\perp+\frac{\Delta_\perp}{2}\right)\cdot x_\perp+i\left(-q_\perp+\frac{\Delta_\perp}{2}\right)\cdot y\perp}\left\langle P+\frac{\Delta_\perp}{2},S\biggl|\Tr [{\cal O}_i(x_\perp){\cal O}'(y_\perp)]\biggr|P-\frac{\Delta_\perp}{2},S\right\rangle \nn
& \qquad \propto - i\frac{S^+}{2P^+}\epsilon_{ij}\left[\left(q_\perp^j +\tfrac{\Delta^j_\perp}{2}\right)f+\left(q^i_\perp-\tfrac{\Delta^j_\perp}{2}\right)g\right]+\cdots.
\end{align}
We observe that in covariant gauges in which the gauge field vanishes at infinity $x^-=\pm \infty$,  both ${\cal O}_i$ and ${\cal O}'$ are gauge invariant (or more properly, BRST invariant). This means that the states ${\cal O}_i|PS\rangle$ and ${\cal O}'|PS\rangle$ are `physical' in that they are annihilated by the BRST operator $Q_B({\cal O}|PS\rangle)=0$ (the Kugo-Ojima condition \cite{Kugo:1979gm}). In much the same way as in the proof of unitarity of the S-matrix in gauge theories, we can insert the intermediate states  
\begin{align}
\left.\sum_X \Tr\left\langle P+\tfrac{\Delta_\perp}{2},S \left|{\cal O}_i(0_\perp)\right|X\right\rangle\left\langle X\left| {\cal O}'(0_\perp)\right|P-\tfrac{\Delta_\perp}{2},S\right\rangle
\right|_{P_\perp^X=-q_\perp}\,,
\end{align}
 and exclude from $X$ the BRST exact states of the form $|X\rangle =Q_B|Y\rangle$. $|X\rangle$ are then gauge invariant states with a positive norm and unit baryon number. A representative of such states is the single nucleon state whose matrix element can be parameterized as 
\begin{align}
\langle P+\tfrac{\Delta_\perp}{2},S|{\cal O}_i(0_\perp)|P^X,S\rangle=\overline{u}\left(P+\tfrac{\Delta_\perp}{2},S\right)\left(a\gamma_i+b \Delta_{\perp i} +cq_{\perp i} \right)u(P^X_\perp=-q_\perp,S)\,.
\end{align}
The structure $\sim \epsilon_{ij}S^+$ comes only from the first term 
\begin{align}
\overline{u}\left(P+\tfrac{\Delta_\perp}{2},S\right)\gamma_i u(P^X_\perp=-q_\perp,S)\approx i\frac{S^+}{P^+}\epsilon_{ij}\left(q_\perp+\tfrac{\Delta_\perp}{2}\right)^j\,, \label{stru}
\end{align}
and this means  $g=0$ for this particular contribution. We cannot extend this argument to the case where $|X\rangle$ is a multiparticle state which consists of one baryon and other hadron species whose  transverse momenta add up to $-q_\perp$.  Yet, it seems reasonable, at least from a naive extrapolation of (\ref{stru}), that the matrix element $\langle \Delta_\perp/2|{\cal O}_i|-q_\perp\rangle$ dominantly depends on the relative transverse momentum between the initial and final states $q_\perp+\Delta_\perp/2$ rather than their sum $-q_\perp+\Delta_\perp/2$. The latter contribution would come from those atypical configurations in which a baryon carries transverse momentum $+q_\perp$ and other hadrons carry $-2q_\perp$ such that their sum is $-q_\perp$.  This indicates  that $|f|\gg |g|$.

\subsection{DGLAP equation}

Finally, we study the double logarithmic limit of the DGLAP equation and directly show that the linear combination $L_g(x)+2\Delta G(x)$ is parametrically  suppressed compared to $\Delta G(x)$. 
Let us assume that $\Delta G(x)$ and $L_g(x)$ are dominant at small-$x$. Then, from (\ref{d1}) and (\ref{d2}) we get
\beq
&&\frac{d}{d\ln Q^2}\Delta G(x)\approx \frac{2C_A\alpha_s}{\pi}\int_x^1\frac{dz}{z}\Delta G(z)\,, \label{glueevolve} \\
&& \frac{d}{d\ln Q^2}L_g(x) \approx \frac{C_A\alpha_s}{\pi} \int_x^1 \frac{dz}{z} (L_g(z)-2\Delta G(z))\,.
\eeq
We see that the linear combination $L_g(x)+2\Delta G(x)$ evolves homogeneously.
\beq
\frac{d}{d\ln Q^2} (L_g(x) +2\Delta G(x)) \approx \frac{C_A\alpha_s}{\pi}\int_x^1 \frac{dz}{z} (L_g(z)+2\Delta G(z))\,. \label{homo}
\eeq
In the double logarithmic limit,  (\ref{homo}) can be solved by 
the standard technique as 
\beq
L_g(x)+2\Delta G(x) \sim \int \frac{dj}{2\pi i} \exp\left(jY + \frac{\xi}{ j}  \right) \sim  e^{2\sqrt{\xi Y}}\,,
\eeq
where $Y=\ln 1/x$ and $\xi \equiv \frac{C_A \alpha_s}{\pi} \ln Q^2$. 
On the other hand, from (\ref{glueevolve}) we get 
\beq
\Delta G(x) \sim e^{2\sqrt{2}\sqrt{\xi Y}} \,.
\eeq
This shows that $|L_g(x)+2\Delta G(x)| \ll |\Delta G(x)|, |L_g(x)|$, as far as $x$-dependence is concerned.

\end{document}